\documentclass[lettersize,journal]{IEEEtran}
\usepackage{amsmath,amsfonts}
\usepackage{algorithmic}
\usepackage{algorithm}
\usepackage{array}
\usepackage[caption=false,font=normalsize,labelfont=sf,textfont=sf]{subfig}
\usepackage{textcomp}
\usepackage{stfloats}
\usepackage{url}
\usepackage{verbatim}
\usepackage{graphicx}
\usepackage{cite}
\usepackage{color}

\usepackage{physics}
\usepackage{multirow}
\usepackage{array}
\usepackage[export]{adjustbox}
\newcolumntype{P}[1]{>{\centering\arraybackslash}p{#1}}
\usepackage{cite}
\begin{document}

\title{Variable Hyperparameterized Gaussian Kernel using Displaced Squeezed Vacuum State}

\author{Vivek Mehta and Utpal Roy\\Indian Institute of Technology Patna, Bihta, Patna 801103, India.}



\maketitle

\begin{abstract}

There are schemes for realizing different types of kernels by quantum states of light. It is particularly interesting to realize the Gaussian kernel due to its wider applicability. A multimode coherent state can generate the Gaussian kernel with a constant value of hyperparameter. This constant hyperparameter has limited the application of the Gaussian kernel when it is applied to complex learning problems. We realize the variable hyperparameterized Gaussian kernel with a multimode-displaced squeezed vacuum state. The learning capacity of this kernel is tested with the support vector machines over some synthesized data sets as well as public benchmark data sets. We establish that the proposed variable hyperparameterized Gaussian kernel offers better accuracy over the constant Gaussian kernel.
\end{abstract}


\section{Introduction}
The kernel methods in machine learning have emerged, particularly during the last decades due to their ability to learn complex patterns over a wide range of data sets \cite{scholkopf2018learning}. Such methods are extensively used due to their ability to learn the decision function of nonlinear separable problems, only by using linear machine learning algorithms without exhausting large amounts of computational resources. A kernel is a positive-definite and symmetric function, whose arguments over the pairs of input data points characterize the similarity of the given pair through their inner product when it is implicitly mapped in the high dimensional space, referred to as feature space.

Applications of machine learning are becoming popular for characterizing quantum phenomena\cite{ahmed2021classification, hsieh2022extract, luo2023detecting, zhang2023entanglement}. Conversely, it is also exciting to see the application of quantum processors for solving machine learning tasks. There are various quantum machine learning models \cite{mishra2021quantum, rebentrost2014quantum, schuld2017implementing, park2020theory, blank2020quantum,blank2022compact, bisarya2020breast}. Schuld highlighted the fact that these quantum machine learning models are equivalent to kernel methods \cite{schuld2021supervised, schuld2019quantum,lloyd2020quantum}. Kernel methods mostly rely on mapping classical inputs to high-dimensional quantum feature space by defining quantum feature mapping, and implicitly accessing these quantum spaces by appropriate quantum measurements. Other than designing quantum learning models, various hybrid machine learning models are also proposed \cite{mitarai2018quantum, schuld2020circuit, havlivcek2019supervised,killoran2019continuous}, where both classical and quantum processors are utilized.

In this paper, we are interested in a particular model where kernels are designed over a quantum processor, referred to as \emph{quantum kernels}. Then a classical processor runs the well-known classical machine learning algorithms, such as support vector machines (SVM), with these quantum kernels. Quantum kernels out of different states of light are reported in the literature \cite{schuld2019quantum,chatterjee2016generalized,li2022quantum, bartkiewicz2020experimental}.
Here, we propose a quantum kernel that is driven from the system of the multimode-displaced squeezed vacuum (DSV) state. This quantum kernel returns a well-known  \emph{Gaussian  kernel} \cite{scholkopf1997comparing, vapnik1998support}:
\begin{equation}
    \label{eq1.1}
    \kappa(\mathbf{x},\mathbf{x}')=exp(-\gamma||\mathbf{x}-\mathbf{x}'||^2),
\end{equation}
where $\mathbf{x}$ and $\mathbf{x}'$ are two inputs, $||.||^2$ is the squared Euclidean distance and $\gamma >0$ is a \emph{variable} hyperparameter. We see explicitly from Eq. (\ref{eq1.1}) that, $\gamma$ decides how much weightage should be given the squared norm of the distance between two inputs, while it implicitly determines the distribution of the input data points onto the feature space and consequently affects the accuracy on the generalization \cite{yang2021parameter, laanaya2011learning}. In other words, it is used to get the balance between variance and bias; and constrains the models from underfitting and overfitting. When the importance of hyperparameter in quantum machine learning is well-understood \cite{shaydulin2022importance, canatar2022bandwidth}, the emphasizes are mostly given to the Gaussian kernel with constant hyperparameter $(\gamma=1)$ by using a quantum state of the multimode coherent state. Such constant hyperparameter limits wider applicability due to a lack of controllability over the design of the kernel and hence the feature mapping. Our proposed quantum kernel with variable hyperparameters will overcome these issues.

The remainder of the paper is organized as follows: in Sec. \ref{sec2}, we discuss the preparation of the kernel with a variable hyperparameter by exploiting the proposed DSV state approach. Section \ref{sec3} will be devoted to the applications of this kernel for learning the nonlinearly separable synthesized data sets as well as public benchmark data sets with the incorporation of SVM. We also highlight the utility of the variable $\gamma$ on the accuracy of learning by comparing it with the constant hyperparameter case for the same data sets. The conclusion and the possible outlook of the work are furnished in the last section \ref{sec4}.

\section{Gaussian  Kernel with Variable Hyperparameter}
\label{sec2}
\begin{figure*}
    \centering
    \includegraphics[width=14cm, height=4.5cm]{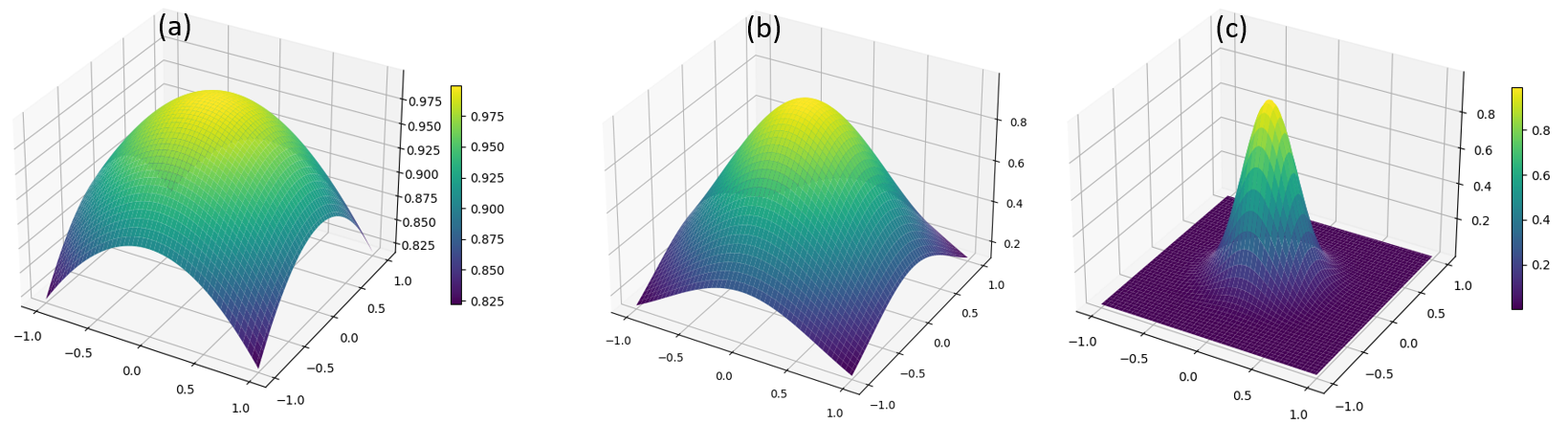}
    \caption{The Gaussian kernel for a pair of scalers, $\kappa(x, x')$, for three different hyperparameters, (a) $\gamma=0.1$, (b) $1$ and (c) $10$, are depicted by fixing one of the scalars to $x=0$ and varying the other, $x'=[-1,1]$, in both the horizontal directions. Plot (b) is the quantum kernel corresponding to a coherent state, which is a special case of our quantum kernel.}
    \label{fig1}
\end{figure*}
The Gaussian kernel is defined when the input data is projected into an infinite-dimensional feature space.  It is a well-known transitional invariant kernel that has wider applications. We aim to implement this kernel using the quantum state of light. We consider our data in $N$-dimensional real spaces, $\mathcal{X}\in \mathbb{R}^N$, and mapped it into the product space of the DSV states, which live in composite space of $N$ infinite-dimensional \emph{dual} Hilbert spaces. The quantum kernel is defined as
\begin{equation}
    \label{eq2.1}
    \kappa(\mathbf{x^p}, \mathbf{x^q})=\prod_{k=1}^N|\langle x^p_k;\eta|x^q_k;\eta\rangle|^2_{\mathcal{H}},\textrm{ } \forall (\mathbf{x^p}, \mathbf{x^q})\in \mathcal{X},
\end{equation}
where $|\langle .|.\rangle|^2_{\mathcal{H}}$ denotes the (square) overlap in quantum Hilbert space, and
\begin{equation}
\label{eq2.2}
    \ket{x;\eta}=\hat{D}(x)\hat{S}(\eta)\ket{0}
\end{equation}
is the DSV state \cite{caves1981quantum, kim1989properties}, obtained by operating  two non-commuting operators: displacement operator $\hat{D}(x)$ \cite{glauber1963coherent} and squeezing operator $\hat{S} (\eta)$ \cite{caves1985new}, over the vacuum state,
\begin{align}
    \hat{D}(x)&=exp(x\hat{a}^\dagger-x^*\hat{a}), \textrm{ and}\notag\\
    \hat{S}(\eta)&=exp\left(\frac{1}{2}(\eta\hat{a}^{\dagger2}-\eta^*\hat{a}^2)\right),
\end{align}
where $x\in\mathbb{R}$ is the coherent displacement variable and the squeezing variable $\eta=re^{i\theta}$ is a complex number: $r, \theta \in \mathbb{R}$. Asterisk denotes the complex conjugate of the number. Both variables have their specific role, such as displacement variables correspond to encoding the vectors, while squeezed variables, which must remain consistent over all product states, act like \emph{variable} hyperparameters. These operations are the functions of bosonic ladder operators, $\hat{a}$ and $\hat{a}^\dagger$, which satisfy the canonical commutative operation, $[\hat{a},\hat{a}^\dagger]=1$.

For the task of computing the inner product, we start by considering the input data as a scaler, $x^p,x^q\in\mathbb{R}$ for which the kernel becomes
\begin{equation}
\label{eq2.3}
    \langle x^p;\eta|x^q;\eta\rangle=\langle0|\hat{S}^\dagger(\eta)\hat{D}^\dagger(x^p)\hat{D}(x^q)\hat{S}(\eta)|0\rangle.
\end{equation}
We need to use the following properties of the displacement operator: $(i)\textrm{ } \hat{D}^\dagger(x)=\hat{D}(-x)$ and $(ii)\textrm{ }\hat{D}(x^p)\hat{D}(x^q)=\hat{D}(x^p+x^q)\text{ }exp\left[\frac{1}{2}(x^qx^{*p}-x^{*q}x^p)\right]$, $[\hat{D}(x^p), \hat{D}(x^q)]\neq0$ \cite{agarwal2012quantum} to write the above equation as
\begin{align}
    \langle x^p;\eta|x^q;\eta\rangle&=\langle0|\hat{S}^\dagger(\eta)\hat{D}(x^q-x^p)\hat{S}(\eta)|0\rangle\times \notag\\
    &\indent\indent\indent exp\left[\frac{-1}{2}(x^qx^{*p}-x^{*q}x^p)\right].\label{eq2.4}
\end{align}
The underlying operator algebra \cite{schumaker1985new,schumaker1986quantum},
\begin{equation*}
    \hat{S}^\dagger(\eta)\hat{D}(x)\hat{S}(\eta)=\hat{D}(\overline{x}),
\end{equation*}
where the argument of the RHS is
\begin{equation}
\label{eq2.5}
    \overline{x}=(x\text{ }coshr+x^*e^{2i\theta}\text{ }sinhr),
\end{equation}
which helps us to obtain the final expression of the kernel in Eq. (\ref{eq2.3}):
\begin{align}
    \langle x^p;\eta|x^q; \eta\rangle&=\langle 0|\hat{D}(\overline{x^q-x^p}) |0\rangle exp\left[-\frac{1}{2}(x^qx^{*p}-x^{*q}x^p)\right]\notag \\
    &=\langle 0|\overline{x^q-x^p}\rangle exp\left[-\frac{1}{2}(x^qx^{*p}-x^{*q}x^p)\right]\notag \\
    &=exp\left(-\frac{1}{2}|\overline{x^q-x^p}|^2\right)\times\notag\\
    &\indent\indent\indent\indent\indent\indent exp\left[-\frac{1}{2}(x^qx^{*p}-x^{*q}x^p)\right]. \label{eq2.6}
\end{align}
It is now necessary to expand the term in the exponent ($|\overline{x^q-x^p}|^2$) of the above final expression with the help of Eq. (\ref{eq2.5}):
\begin{align}
    |\overline{x^q-x^p}|^2&=\left(\overline{x^q-x^p}\right)^*\text{ }\overline{x^q-x^p}\notag\\
    &=|x^q-x^p|^2(cosh^2r+sinh^2r) + \notag \\
    &\indent\big[((x^q-x^p)^*)^2 e^{-2i\theta}+\notag\\
    &\indent\indent\indent(x^q-x^p)^2 e^{2i\theta}\big]coshr\text{ }sinhr. \label{eq2.7}
\end{align}
For real $x^p\textrm{ and }x^q$, Eq. (\ref{eq2.6}) will reduce to
\begin{align}
    \langle x^p;\eta|x^q; \eta\rangle=exp\bigg(-\frac{1}{2}(x^q-x^p)^2\times \notag\\
    (cosh^2r+sinh^2r+2\text{ }cos2\theta\text{ } coshr \text{ }sinhr)\bigg). \label{eq2.8}
\end{align}
Here, we explore two interesting cases, based on two distinct phases of the squeezing parameter. These choices provide the expressibility of the hyperparameter.
\begin{equation}
\label{eq2.9}
    \langle x^p;\eta|x^q; \eta\rangle=\begin{cases}
        exp\left(-\frac{e^{r}}{2}(x^q-x^p)^2\right), & \text{if }\theta=0 \\
         exp\left(-\frac{e^{-r}}{2}(x^q-x^p)^2\right), & \text{if }\theta=\frac{\pi}{2} .
    \end{cases}
\end{equation}
An appropriate phase should be chosen from these expressions, corresponding to the hyperparameter value that lies below 1, and otherwise. The kernel corresponding to the single-valued real input-data can be easily generalized to a vector input-data, and hence the quantum kernel in Eq. (\ref{eq2.1}) will become
\begin{align}
    \kappa(\mathbf{x^p}, \mathbf{x^q}) &=\prod_{k=1}^N\langle x^q_k;\eta|x^p_k; \eta\rangle\langle x^p_k;\eta|x^q_k; \eta\rangle, \notag \\
    &=exp( -\gamma||\mathbf{x^p}- \mathbf{x^q}||^2), \label{eq2.10}
\end{align}
where $||.||^2$ is squared Euclidean distance and $\gamma$ is the hyperparameter, which can be taken by suitably choosing the correct squeezing variable: $\gamma=e^r \text{ or } e^{-r}$. The above equation is the \emph{exact} realization of a Gaussian kernel with variable hyperparameter $\gamma$, which determines the width of the kernel. A Gaussian kernel with variable hyperparameter is shown in Fig. (\ref{fig1}) for three different hyperparameters. The Gaussian becomes narrower with increasing $\gamma$. A special case for $\gamma=1$ gives us the quantum kernel corresponding to a coherent state.

\textbf{\textit{Optical Circuital Implementation of the Kernel}}
It is worthwhile to discuss the method of realization of our proposed kernel using real devices. We find it feasible to implement it in an optical computational platform through the integrated programmable photonic computers, which have been established as a mature hardware platform for demonstrating various computational and quantum technological tasks.
\begin{figure*}
    \centering
    \includegraphics[width=9.5cm, height=4cm]{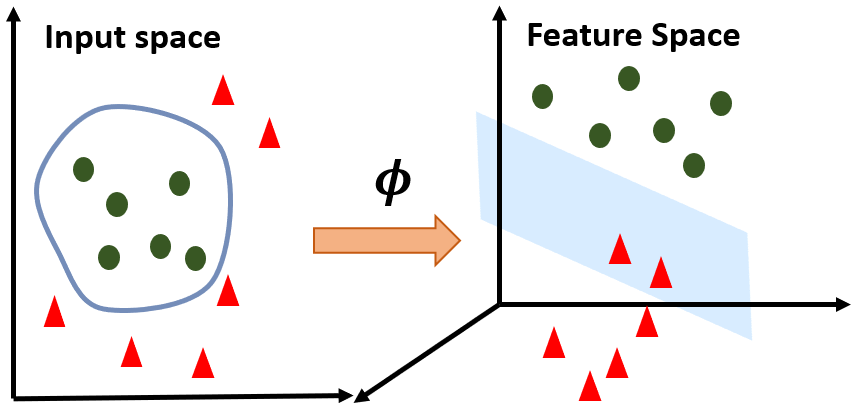}
    \caption{A kernel implicitly transforms a nonlinear problem into a linear separable problem, which is equivalent to the nonlinear mapping of the input space to the feature space, by appropriate feature space $\phi$, where data points belonging to the different classes are separable by the appropriate hyperplanes.}
    \label{fig2}
\end{figure*}

All the required quantum mechanical operations for deriving the mathematical expression of our kernel can be realized by the linear optical elements, such as beam splitters, phase shifters, and the nonlinear optical elements like parametric amplifiers (or squeezers) \cite{braunstein2005squeezing}. The circuital realization of the kernel for the given scalar inputs, $x^p,x^q\in\mathbb{R}$, can be easily generalized to an arbitrary dimensional input space. The first stage will be to initialize a qmode with the input vacuum state, which will be followed by a sequence of the gate operations to result into the quantum state,
\begin{equation}\label{eq2.1.1}
     \ket{\psi}=\hat{S}^\dagger(\eta)\hat{D}^\dagger(x^p)\hat{D}(x^q)\hat{S}(\eta)|0\rangle.
\end{equation}
Further, one needs to apply a nonlinear photodetection process, mathematically represented by the projection operator, $\ket{0}\bra{0}$ on the state in Eq. (\ref{eq2.1.1}). The probability of detecting no photon in the qmode is eventually the proposed kernel as seen in Eq. (\ref{eq2.10}),
\begin{align}
    \langle\psi|0\rangle\langle0|\psi\rangle&=\langle x^q;\eta|x^p; \eta\rangle\langle x^p;\eta|x^q; \eta\rangle\notag\\
    &= |\langle x^p;\eta|x^q; \eta\rangle|^2.
\end{align}
The circuit should be operated for obtaining the probability and to estimate the value our the kernel for a given pair of scalars.

\section{Demonstration and Classification Accuracy on Learning with the DSV State Quantum Kernel:}
\label{sec3}

\begin{figure*}
\centering
    \includegraphics[width=14cm, height=13cm]{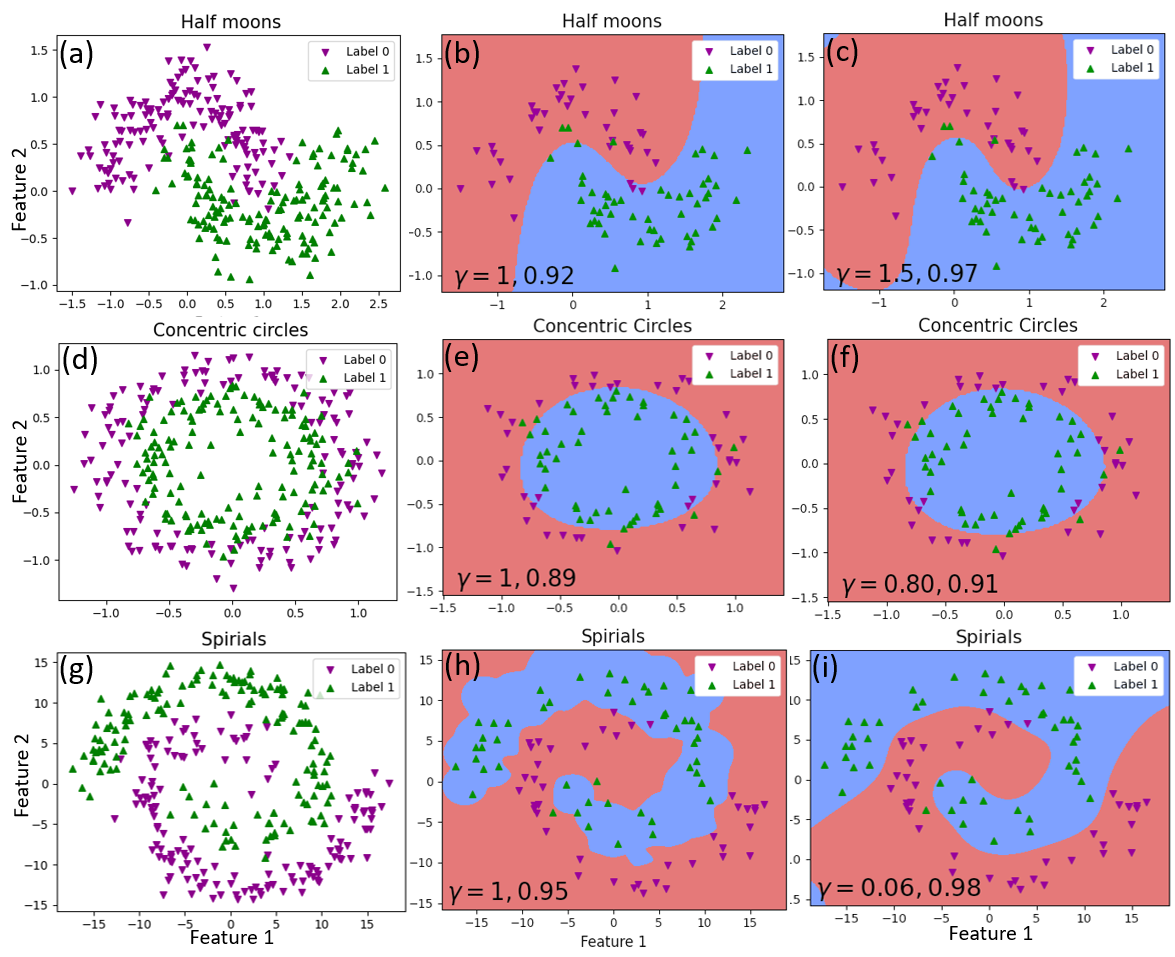}
    \caption{Half moons, concentric circles, and spirals are three synthesized data sets, shown in plots (a), (d), and (g), respectively. Plots (b), (e) and (h) are the decision boundary plots corresponding to the coherent state-based quantum kernel ($\gamma=1$). The value of the hyperparameter and classification accuracies are shown in the lower-left part of these plots. Similarly, the decision plots for our DSV state-based quantum kernel with variable hyperparameters are depicted in (c), (f) and (i) for $\gamma=1.5$, $\gamma=0.8$ and $\gamma=0.06$, respectively, along with the corresponding values of hyperparameter and classification accuracies in the lower-left corner. In all the decision plots, only test data points are shown.}
    \label{fig3}
\end{figure*}

We first describe the SVM \cite{boser1992training,scholkopf1999advances}, which offers a favorable manifestation of the fusion between a statistical idea and a working practical application \cite{hearst1998support}. It belongs to the class of hyperplane learning algorithms. With the conjunction of the kernels, the learning ability of the SVM  can be extended to nonlinear separable problems in input space without increasing computational resources (see Fig. (\ref{fig2})). A nonlinear separable complex learning problem can be solved by the linear machine learning algorithms like SVM with the assistance of a kernel. The SVM can find an optimal hyperplane among all separating hyperplanes in the feature space, that has the maximum margin of separation between any training data points and the hyperplane. The optimal hyperplane for the linear separable classification problem in the inner product space $\mathcal{F}$ has the maximum margin from each class by the support of a small subset of the training data set, which are referred to as support vectors. These supporting data points are closest to the maximum marginal hyperplane. More insight about the feature mapping and kernels can be drawn by the following theorem. \\
\textbf{Theorem}: Given an input space, $\mathcal{X}$, and a nonlinear mapping function, $\phi:x\in \mathcal{X}\rightarrow \phi(x)\in\mathcal{F}$, the inner products in the feature space $\mathcal{F}$ are represented by
\begin{equation}
\label{eq3.1}
    \kappa(x^p,x^q)=\langle\phi(x^p),\phi(x^q)\rangle_{\mathcal{F}}
\end{equation}
for any pairs $x^p,x^q\in\mathcal{X}$ and it defines a positive-definite and symmetric function as kernel. It is worth to be highlighted that, $\langle.,.\rangle_{\mathcal{F}}$ signifies a inner product in the feature space and this space is also a Hilbert space, unlike a dual Hilbert space. Secondly, there is no constraint over the structure of input spaces and it will consist of graphs, binary strings, matrices, \emph{etc}.

\begin{figure*}
\centering
    \includegraphics[width=14cm, height=9cm]{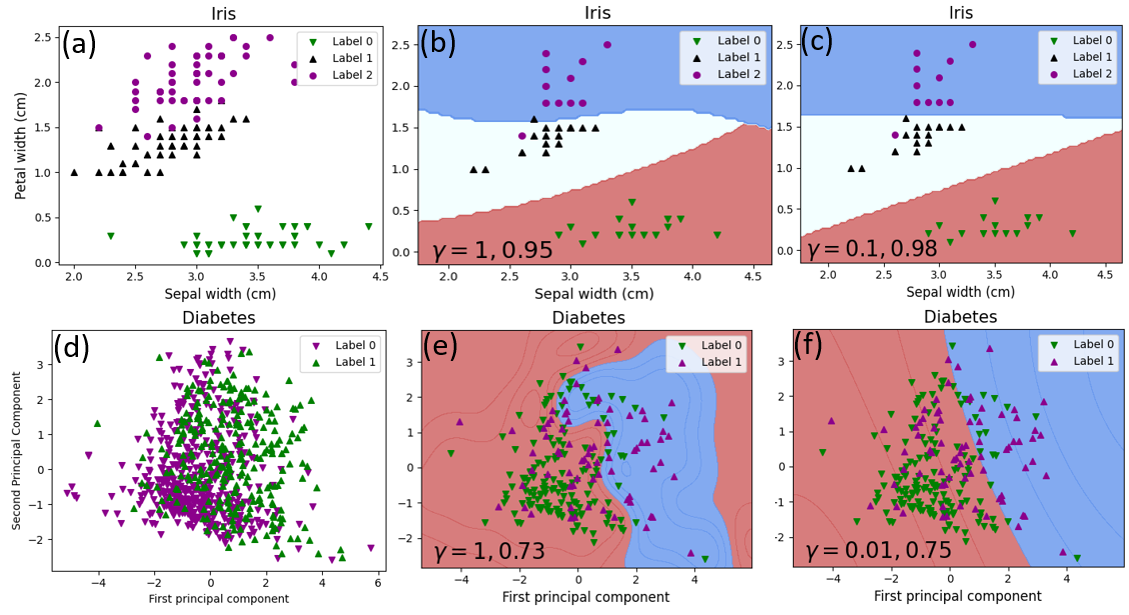}
    \caption{(a) the distribution of the data points when we choose two features of the Iris data set, sepal width, and petal width. Plots (b) and (c) are the decision boundary plots along with test points distribution, respectively, when coherent state- and our DSV state-based quantum kernels are being deployed for classification. The hyperparameters and corresponding classification accuracies are shown in the lower-left parts. The distribution of the diabetes data points in (d) is manifested for coherent state- and DSV state-based kernels in (e) and (f), respectively.}
    \label{fig4}
\end{figure*}
Let us now formulate a supervised learning problem. Consider a training data set consisting of $M$ vectors, $\mathcal{D}=\{\mathbf{x}^m,y^m\}_{m=1}^M\subset \mathbb{R}^N\times\{1,\dots, L\}$, live in $N$-dimensional real space and their labels belong to any one of the $L$ classes. Particularly, we emphasize on a binary classification problem without any loss of generalization. In \emph{direct }space, any separating hyperplane in the feature space is
\begin{equation}
\label{eq3.2}
    \langle\mathbf{w},\phi(\mathbf{x})\rangle_{\mathcal{F}}+w_0=0,
\end{equation}
and the corresponding decision function is
\begin{equation}
    f(\mathbf{x})=sgn\left(\langle\mathbf{w},\phi(\mathbf{x})\rangle_{\mathcal{F}}+w_0\right),
\end{equation}
where $w_0\in \mathbb{R} \text{ and } \mathbf{w}\in \mathbb{R}^N$ are a bias and a learning vector, respectively. These are required to be learned during the training phase with the training data set. Here, $sgn (\dots)$ is a sign function and $\mathbf{x}$ is the new unseen data. While in the \emph{dual} space, the decision function becomes
\begin{equation}
\label{eq3.3}
   f(\mathbf{x})=sgn\left( \sum_{m=1}^M\alpha_m y^m\kappa(\mathbf{x}^m,\mathbf{x})+w_0\right),
\end{equation}
where $\{\alpha_m\}$ are called dual parameters and $\kappa(.,.)$ is a kernel, defined in the above theorem. Solving the following optimization problem is needed to get a maximum marginal separating hyperplane:
\begin{align}
\underset{\mathbf{w}, b}{min} \quad ||\mathbf{w}||^2
\label{eq3.4}
\end{align}
subject to $y^m\left(\langle\mathbf{w},\phi(\mathbf{x}^m)\rangle_{\mathcal{F}}+w_0\right)\geq1$ for $m=1,\dots,M$. This optimization problem can also be transformed into the dual space employing the Lagrange and then the dual-optimization problem becomes
\begin{align}
\underset{\mathbf{\alpha}}{max}\textbf{ }L(\mathbf{\alpha})=\sum_{m=1}^M\alpha_m-\frac{1}{2}\sum_{p,q=1}^M\alpha_p\alpha_q\kappa(\mathbf{x}^p,\mathbf{x}^q) \label{eq3.5}
\end{align}
subject to $\alpha_m\geq0\text{ for }m=1,\dots,M $ and
\begin{align}
     \sum_{m=1}^M\alpha_m y^m=0.\notag
\end{align}
When we assign the above objective function to the quadratic programming software, we come up with the optimized value of the dual parameter vector, $\mathbf{\alpha}^*$. According to Karaush-Kuhn-Tucker's condition, dual parameters corresponding to the supporting vector (SV) are only non-zero. Using $\mathbf{\alpha}^*$, we get the optimized direct parameter vector as $\mathbf{w}^*=\sum_{\phi(\mathbf{x}^m)\in SV}\alpha^*_my^m\phi(\mathbf{x}^m)$ and the optimized value of bias, $w_0^*$. This equation, $\left(\langle\mathbf{w}^*,\phi(\mathbf{x}^m)\rangle_{\mathcal{F}}+w_0\right)=1$, has to be used for any  supported vector, $\mathbf{x}^m$. Upon achieving the optimal learning, the decision function in Eq. (\ref{eq3.3}) needs to be used to label the unseen data $\mathbf{x}$.

After getting familiar with the SVM and its integration with the kernel, it is important to see the implication of the Gaussian kernel with variable hyperparameter with respect to its constant version in terms of classification accuracy over the synthesis data sets (see Fig. \ref{fig3}) as well as the public benchmark data sets ( see Fig. \ref{fig4}). Each synthesis data set consists of $300$ data points and each label takes equal halves of data points. On the other hand, we take two public benchmark data sets: Iris and Diabetics. Iris consists of 150 samples, that belong to three labels and each sample has four features. The diabetes data set consists of 768 binary labeled samples and each has eight numerical features. Since the number of features is high and therefore, we employ the principle component analysis which comes with the most correlated two features: the first principle component and the second principle component. Here, we use Scikit-learn, which is a Python-based machine learning library to get the classification accuracy. Each data set is disintegrated into $70:30$ ratios for training and testing purposes, respectively. 

Some synthesized data sets, such as half moons, concentric circles, and spirals are taken for the first stage of demonstration and decision boundaries are depicted in Fig. \ref{fig3}. When Fig. \ref{fig3}(a), (d), and (g), respectively, manifest the data, the corresponding decision boundaries for a coherent state-based kernel ($\gamma=1$) are shown in Fig. \ref{fig3}(b), (e) and (h). The value of the hyperparameter and classification accuracies are shown in the lower-left part of these plots in sequence. If we compare these accuracies against that of our DSV state-based quantum kernel with variable hyperparameters (depicted in Fig. \ref{fig3}(c), (f) and (i)), $\gamma=1.5$, $\gamma=0.8$ and $\gamma=0.06$, respectively, we observe a clear improvement. On generalization, we find that classification accuracy is improved for each of the data sets when variable hyperparameter $\gamma$ is used during training, other than $1$. The merit of our kernel method with variable hyperparameter over the constant hyperparameter methods is also verified through two public benchmark data sets: Iris and Diabetics, and delineated in Fig. \ref{fig4}. The sepal width-petal width and first-second principal components distribution plots finds their decision boundaries with higher accuracy for our presented kernel, when compared to the coherent state-based kernel methods.

Hence, the learning capacity of this proposed kernel with variable hyperparameter is established to provide a generalized classification scheme, where the variable hyperparameterized Gaussian kernel offers better accuracy over the constant Gaussian kernel.

\section{Conclusions}
\label{sec4}
We proposed the Gaussian kernel with the variable hyperparameter using the DSV state, where the squeezing parameter is linked to the variable hyperparameter, while the coherent state parameters are used to encode the entries of the inputs. The mathematical steps for the derivation of this quantum kernel are provided by exploiting quantum optical operator algebra. The support vector machine algorithm is used to find the optimal hyperplanes in the feature space. We have tested the protocol over diverse data sets, belonging to the synthesized data sets and also the public benchmark data sets. The proposed quantum kernel provides the design, that overwhelms the limitation to work with a coherent state-based quantum kernel. Moreover, the underlying classification accuracies over all the data sets establish that, a Gaussian kernel with variable hyperparameter offers a better design of the hybrid quantum-classical model. The scheme can be further applied for other useful data sets of need and a variety of hyperparameter values.


\end{document}